\documentclass[10pt, aps, pra, superscriptaddress,
               nofootinbib, 
               amsmath, amssymb, twocolumn,
               preprintnumbers, showpacs,
               raggedbottom,
               floatfix]{revtex4-1}

\newcommand{\figdir}{./data/_generated/figures}  % Where are the figures?
\usepackage[arXiv, todo]{my_paper}        % Use this for the arXiv

\let\email\exclude                        % Don't show email addresses

\usepackage{my_acronyms}                  % acronyms for this paper.

\usepackage[utf8]{inputenc}

\newcommand{\cSR}{c_{\chi}}

\renewcommand{\todo}[1]{}
\let\Re\relax

\DeclareMathOperator{\Re}{Re}

\begin{document}
\title{Validating Simple Dynamical Simulations of the Unitary Fermi Gas}

\author{Michael McNeil Forbes}
\affiliation{Institute for Nuclear Theory, University of Washington,
  Seattle, Washington 98195--1550 \mysc{usa}}
\affiliation{Department of Physics, University of Washington, Seattle,
  Washington 98195--1560 \mysc{usa}}
\affiliation{Department of Physics \& Astronomy, Washington State University,
  Pullman, Washington 99164--2814, USA}
\email{mforbes@alum.mit.edu}

\author{Rishi Sharma}
\affiliation{T\textsc{riumf}, Vancouver,  British Columbia, V6T 2A3, Canada}
\affiliation{Department of Theoretical Physics, Tata Institute of Fundamental
  Research, Homi Bhabha Road, Mumbai 400005, India}
\email{rishi@triumf.ca}

\date{\today}

\begin{abstract}
  \noindent
  We present a comparison between simulated dynamics of the unitary fermion gas
  using the \gls{SLDA} and a simplified bosonic model, the \gls{ETF} with a
  unitary equation of state. Small amplitude fluctuations have similar dynamics
  in both theories for frequencies far below the pair breaking threshold and
  wave vectors much smaller than the Fermi momentum, and the low frequency
  linear responses match well for surprisingly large wave vectors, even up to
  the Fermi momentum.  For non-linear dynamics such as vortex generation, the
  \gls{ETF} provides a semi-quantitative description of \gls{SLDA} dynamics as
  long as the fluctuations do not have significant power near the pair breaking
  threshold, otherwise the dynamics of the \gls{ETF} cannot be trusted.
  Nonlinearities in the \gls{ETF} tends to generate high-frequency fluctuations,
  and with no normal component to remove this energy from the superfluid,
  features like vortex lattices cannot relax and crystallize as they do in the
  \gls{SLDA}.  We present a heuristic diagnostic for validating the
  reliability of \gls{ETF} dynamics by considering the approximate conservation
  of square of the gap: $\int \abs{\Delta}^2$.
\end{abstract}
\preprint{\mysc{INT-PUB-13-033}}

\setcounter{page}{0}
\clearpage

\maketitle
\glsresetall

%\section{Introduction}\noindent
\lettrine{T}{he motion} of cold Fermi superfluids under dynamical stimuli has
been of interest to a variety of research fields. A classic example is the
discovery of the Josephson effect~\cite{Josephson:1974} in superconductors. Now
it is possible to track the motion of magnetic vortices in real
time~\cite{Park:2003,Zhu:2005,Choe:2004,Compton:2006}, and to ramp the fermionic
interaction in cold-atom atomic experiments from the \gls{BCS} to the \gls{BEC}
regime~\cite{Ketterle:2008}. Furthermore, by changing the trapping potential,
phenomena like particle transport~\cite{2005:Bloch,Schneider:2012}, and cloud
collision dynamics~\cite{Joseph:2011} have been quantitatively measured. More
recently, ``heavy solitons'' were observed oscillating in elongated
traps~\cite{Yefsah:2013} with very long periods. Many nuclear
responses~\cite{Stetcu:2011} and reaction processes~\cite{Bulgac:2010xf} are
manifestation of collective dynamics of nucleons, and vortex pinning and
unpinning likely plays a role in generating glitches in the spin down of neutron
stars~\cite{Link:1999}.

Despite this diverse interest, simulating fermionic \gls{QHD} -- even with
simplified \gls{TDDFT} models -- remains a computational challenge, requiring
world-class computing resources for even relatively simple
problems~\cite{Bulgac:2011b}.  Direct simulation of many macroscopic phenomena
lies outside the realm of current technology, so in this paper, we validate to
what extent a computationally simple model called the \gls{ETF} model can
characterize the dynamics of the strongly interacting \gls{UFG}, finding that it
performs well for low-frequency dynamics, and identifying its limitations.  This
validation played a crucial role in solving the mystery of the ``heavy
solitons'' observed in~\cite{Yefsah:2013} where the \gls{ETF} was used to
demonstrate the consistency of the observations with vortex rings instead of
solitons~\cite{Bulgac:2013d}.

The computational problem is that most fermionic superfluid \glspl{TDDFT} (of
the Kohn-Sham variety such as the \gls{BdG} mean-field equations or \gls{HFB}
equations) require evolving hundreds of thousands of single-particle
wavefunctions, occupying vast amounts of memory.  In contrast, the
\gls{GPE}~\cite{Fetter:1998Review,PS:2002} provides an attractive computational
method for studying bosonic superfluids where the superfluid is represented by a
single wavefunction for the condensed state.  The \gls{ETF} model considered
here has the same computational simplicity, and thus can be applied to macroscopic
systems. We find that it performs well for low-frequency dynamics, suggesting
that it might provide the basis for a practical method of simulating macroscopic
volumes of fermionic superfluids required to understand phenomena like neutron
star glitches.

As expected, the \gls{ETF} model fails when significant energy is present near
the pair-breaking scale. In fermionic \glspl{TDDFT}, energy above the pair
breaking threshold converts superfluid to normal fluid, hence we suggest that
the conservation of $\int\abs{\Delta}^2$ might be a good measure of the
applicability of the \gls{ETF} and validate this diagnostic.  We also suggest
how the model might be improved to better model more energetic scenarios.

The \gls{UFG} is a universal model for dilute Fermi gasses comprising two
species of the same mass interacting with a zero-range resonant attractive
interaction of infinite $s$-wave scattering length.  It provides an ideal
problem to benchmark many-body techniques for several reasons: it has a simple
and universal \gls{EoS} but remains highly non-perturbative with strong
interactions, is directly realized in cold-atom experiments~\cite{Zwerger:2011},
and provides a good approximation for the dilute neutron
matter~\cite{Carlson:2012} in the crusts of neutron stars.  This universal
system is stable, and the absence of a length scale for the interaction implies
that the energy-density $\mathcal{E}(\rho) = \xi\mathcal{E}_{FG}(\rho)$ is
characterized by the single universal dimensionless coefficient $\xi$ known as
the Bertsch parameter~\cite{Bertsch:1999:mbx}. (Here $\mathcal{E}_{FG}(\rho) =
\frac{5}{3}\rho E_F(\rho)$ is the energy-density of the non-interacting
system with total density $\rho = \rho_a + \rho_b = k_F^3/3\pi^2$,
$E_F(\rho) = \hbar^2 k_F^2/2m$ is the Fermi energy, and $k_F$ is the
Fermi wave-vector.)  Despite the simple form of the \gls{EoS}, the system is
strongly interacting and admits no perturbative expansions. Significant effort
has been put into determining the Bertsch parameter $\xi$ over the past decade,
and only recently has it been computed~\cite{CCPS:2003, ABCG:2004, Bulgac:2006a,
  Lee:2008, FGG:2010, Forbes:2012} and measured~\cite{Sylvain-Nascimbene:2009lr,
  Horikoshi:2010, Ku:2011} to high precision (see~\cite{Endres:2012} for a
survey).  The current best fit value $\xi=0.3742(5)$ is obtained by consistently
fitting both \gls{QMC} and experimental results~\cite{Forbes:2012} with a
self-consistent fermionic \gls{DFT} called the \gls{SLDA}.

\glsunset{TDSLDA} The time-dependent generalization of the \gls{SLDA}
(\gls{TDSLDA}) provides a model for directly studying time-dependent phenomena
in the \gls{UFG} (see~\cite{Bulgac:2011a, *Bulgac:2011} for a review).  The
\gls{TDSLDA} models both superfluid hydrodynamics and normal hydrodynamics,
including pair breaking effects, the superfluid-normal transition, and
finite-size (shell) effects.  Many different dynamical processes have been
described in~\cite{Bulgac:2011b}, including vortex nucleation through stirring,
vortex-vortex interactions, vortex ring formation etc.  These simulations,
however, required super-computing resources for even modest physical
volumes. The largest system studied in~\cite{Bulgac:2011b} contained $\sim
\num{500}$ particles represented by \num{70000} wavefunctions on a $32\times
32\times 192$ lattice.  To compare, typical cold-atom experiments comprise some
$10^5$ particles~\cite{KZ:2008_full}, which would severely tax current
computational resources, even if symmetries are utilized. Similarly, while the
dynamics of a single vortex in neutron matter may be within reach of cutting
edge computing~\cite{Bulgac:2013a}, simulating multiple vortices separated by
several lattice lengths will require significantly more resources.  Thus,
validating and generalizing computationally more efficient methods like the
\gls{ETF} model is critical for scaling calculations up to macroscopic systems.

The \gls{ETF}~\cite{Kim:2004a, *Kim:2004, Salasnich:2008b, *Salasnich:2008,
  *Salasnich:2008E} is essentially a bosonic theory describing
dimers/Cooper-pairs in the \gls{UFG} with a single collective condensate
wavefunction that has been used to analyze the expansion and breathing mode
frequencies of cold atomic gases in a trap~\cite{Kim:2004a, *Kim:2004,
  *Kim:2005, *Kim:2005a}, their surface oscillations~\cite{Salasnich:2010,
  *Salasnich:2012}, collisions of clouds of fermions~\cite{Ancilotto:2012,
  *Ancilotto:2012a}, vortex generation~\cite{Ancilotto:2013}, vortex
pinning~\cite{Bulgac:2013a}, instabilities~\cite{Gautam:2013}, and soliton
dynamics~\cite{Khan:2013}. While the \gls{ETF} has the same symmetries, and can
be tuned to have the same \gls{EoS} as the full theory, one expects poor
behaviour when excitations approach the
pair-breaking threshold set by the gap $\hbar\omega > 2\Delta \approx E_F$. The
only low energy degree of freedom -- the superfluid phonon -- exists in both
theories, and matching the \glspl{EoS} ensures that speed of sound is the same.
This ensures that the linear response for small frequencies and momenta match,
but we find good agreement for small frequencies even at finite momenta $q
\sim k_F$, suggesting that the \gls{ETF} could be a good description
of~\gls{SLDA} dynamics for slowly varying probes.  Indeed, the \gls{ETF} seems
to do a good job of describing bulk dynamics in regimes where pair-breaking
effects play a minor role, but exhibits notable departures as one introduces
excitations near the pair-breaking threshold.  We verify this behaviour by
comparing with existing fermionic~\cite{Bulgac:2011b} simulations and find
certain diagnostics to check whether the bosonic simulation can be expected to
be a good description of the fermionic problem.

In Sections~\ref{sec:modelSLDA} and~\ref{sec:model} we review the \gls{TDSLDA}
and the \gls{ETF} models. We compare the linear response for time independent
fluctuations in Section~\ref{sec:static-response} and for dynamic fluctuations
in Section~\ref{sec:linear-response}. In Section~\ref{sec:nonlinear-response} we
compare the \gls{ETF} with \gls{TDSLDA} dynamics for a family of simulations
where vortices are created and nonlinear effects are important and conclude in
Section~\ref{sec:conclusion}. We present a discussion of the numerical
implementation in Appendix~\ref{sec:numer-impl} and give some details of the
simulation parameters in Appendix~\ref{sec:trap-params}.

\section{The SLDA}\label{sec:modelSLDA}\noindent
We start with a brief review of the \gls{SLDA} \gls{DFT}.
\glsreset{DFT}\Gls{DFT} is in principle an exact approach, widely used in
nuclear physics (see~\cite{Drut:2010kx} for a review) and in quantum chemistry
to describe normal (i.e., non-superfluid) systems.  It provides a framework
capable of assimilating \textit{ab initio} and experimental results into a
computationally tractable and predictive framework.  The original Hohenberg-Kohn
formulation~\cite{HK:1964} proves the existence of an energy functional
$E[\rho(x)]$ such that the density $\rho_0(x)$ and energy $E_0$ of the ground
state of an interacting system in an external potential $V(x)$ can be found by
minimizing
\begin{gather}
  E_{0} = 
  \min_{\rho(x)} \left(E_{\textsc{dft}}[\rho(x)] + \int\d^3{x}\;V(x)\rho(x)\right).
\end{gather}
Dynamics may be described by an extension commonly referred to as
\glsreset{TDDFT}\gls{TDDFT}~\cite{Rajagopal:1973rt, *Peuckert:1978fr,
  *Runge:1984mz} that describes the evolution of the one-body number density in
the presence of an arbitrary one-body external field.  As in the static case one
can prove the existence of a functional from which one can determine the exact
time-dependent number density for a given quantum system~\cite{Rajagopal:1973rt,
  *Peuckert:1978fr, *Runge:1984mz}.  Unfortunately, these theorems do not
specify the form of the functional $E_{\textsc{dft}}[\rho(x)]$.

Instead, one must rely on physically motivated models and benchmark them.  To
model the \gls{UFG}, a simple form known as the
\glsreset{SLDA}\gls{SLDA}~\cite{Bulgac:2002uq, *BY:2002fk} has been successfully
benchmarked against \textit{ab initio} \gls{QMC} calculations~\cite{FGG:2010,
  *Forbes:2012, *Forbes:2012a}.  It is a local functional of the density $\rho$
and two additional densities: a kinetic density $\tau(x) \propto
\braket{\vect{\nabla}\psi^\dagger\cdot\vect{\nabla}\psi}$ (following the
Kohn-Sham formulation~\cite{Kohn:1965fk}) which is required to model finite-size
(shell) effects, and an anomalous density $\nu \propto \braket{\psi\psi}$
required to model pairing effects (see~\cite{Bulgac:2007a, Bulgac:2011a,
  *Bulgac:2011, FGG:2010, *Forbes:2012, *Forbes:2012a} for a discussion).  The
resulting \gls{SLDA} energy density functional
\begin{gather}
  \mathcal{E}_{\textsc{slda}} =
  \frac{\hbar^2}{m}\left(
    \frac{\alpha}{2}\tau_{+} +  g\nu^{\dagger}\nu\right) 
  + \beta \mathcal{E}_{FG}(\rho) \nonumber\\
  g^{-1} = \frac{\rho^{1/3}}{\gamma} - \frac{k_c}{2\pi\alpha}
  \label{eq:DF_SLDA}
\end{gather}
has three dimensionless parameters: an inverse effective mass $\alpha$, a
self-energy $\beta$, and a pairing parameter $\gamma$.  (The anomalous density
$\nu$ diverges in the local approximation requiring regulation expressed through
a cutoff $k_c\rightarrow \infty$.)  One typically solves the \gls{SLDA} for
homogeneous matter, expressing $\beta$ and $\gamma$ in terms of the physically
relevant Bertsch parameter $\xi$ and the $T=0$ pairing gap $\eta = \Delta/E_F$.
If the effective mass parameter $\alpha \neq 1$, then one must also introduce a
term involving currents to restore Galilean covariance that slightly complicates
the numerical implementation.  For this reason, and since $\alpha \approx 1$,
the \gls{TDSLDA} employed in practice typically sets $\alpha = 1$
\cite{Bulgac:2011b} and we shall compare to these results in this paper.

To work with this \gls{DFT}, one expresses $\tau$, $\rho$, and $\nu$ in terms
of a set of single-particle orbitals that obey a set of self-consistency
equations similar to the \gls{BdG} mean-field equations
\begin{gather}
  \I \hbar \frac{\partial}{\partial t}
  \begin{pmatrix}
    u_n(x,t) \\ 
    v_n(x,t)
  \end{pmatrix} = 
  \begin{pmatrix}
    \op{K} + U & \Delta \\
    -\Delta^* &  -\op{K} - U
  \end{pmatrix}
  \begin{pmatrix}
    u_n(x,t) \\ 
    v_n(x,t)
  \end{pmatrix}
  \label{eq:TDSLDA}
\end{gather}
where $U[\tau, \rho, \nu]$ and $\Delta[\tau, \rho, \nu]$ are functions of the
densities obtained by minimizing~\eqref{eq:DF_SLDA}. The computational
difficulty is that one must simultaneously evolve many single-particle
wavefunctions, $(u_n, v_n)$, and one eventually becomes limited by memory
(\num{70000} wavefunctions on a $32\times 32\times 192$ grid requires
\SI{200}{GB} for a single step).

We note that the \gls{SLDA} reproduces the variational \gls{BdG} mean-field
equations if one sets the effective mass to unity $\alpha=1$, removes the
self-energy $\beta = 0$, and tunes the pairing interaction with the usual
pseudo-potential $\gamma^{-1} = 0$.  This well-studied approximation captures
the same qualitative physics as the \gls{SLDA}, but does not provide a reliable
quantitative picture (the lack of a self-energy $\beta=0$ for example
incorrectly predicts a non-interacting polaron with zero binding energy).  The
\gls{SLDA} will reproduce this model if one fixes the parameters $\alpha=1$,
$\xi=0.5906\cdots$ and $\eta=0.6864\cdots$.  Since this \gls{BdG} model has been
widely studied, we include a comparison between it and the \gls{ETF} tuned to
this ``incorrect'' value of $\xi$ along with the comparison to the \gls{SLDA}.

\section{The ETF Model}\label{sec:model}\noindent
In contrast, the \glsreset{GPE}\gls{GPE}~\cite{Fetter:1998Review,PS:2002}
commonly used to model bosonic superfluids, requires storing and evolve only a
single complex wavefunction $\psi$ representing the condensate, thereby allowing
one to explore significantly larger systems.  To apply this approach to the
\gls{UFG} we note that one can describe the \gls{BEC} limit of strong attraction
as a Bose gas of dimers.  Hence, we introduce $\Psi(x,t)$ as the collective
dimer wavefunction into a modified \gls{GPE}~\cite{Salasnich:2008b,
  *Salasnich:2008, *Salasnich:2008E}:\todo{Properly address history of ETF
  model}
\begin{subequations}\label{eq:ETF}
  \begin{gather}
    E_{\textsc{etf}}[\Psi] = 
    \intd{3}{\vx}\left(
      \frac{\abs{\nabla\Psi(\vx)}^2}{4m}
      + V(\vx)\rho + g(\rho)
    \right),\label{eq:E_UFG}\\
    \I\hbar\partial_t\Psi = \mat{H}\Psi = 
    \left(
      -\frac{\hbar^2\nabla^2}{4m} + 2[V + g']
    \right)\Psi,\\
    \begin{aligned}
      \rho &= 2\abs{\Psi}^2, &
      g(\rho) &= \xi\mathcal{E}_{FG}(\rho), &
      g'(\rho) &= \xi E_F(\rho).
    \end{aligned}
  \end{gather}
\end{subequations}
One should think of this as a \gls{GPE} for the ``dimer'' Cooper pairs. The
bosonic dimers are described by the collective wavefunction $\Psi(x,t)$ with the
interpretation that $\abs{\Psi}^2$ is the dimer density, hence the total density
$\rho = 2\abs{\Psi}^2$ has a factor of $2$.  Likewise, the bosonic mass $m_B =
2m$ is twice the fermionic mass, accounting for the factor of $4m = 2m_B$ in the
kinetic term.  This picture becomes more accurate as the dimers become more
tightly bound toward the \gls{BEC} regime where a \gls{GPE} description of the
bosonic dimers is applicable.  Finally, whereas the \gls{GPE} has a quartic
self-interaction related to the dimer-dimer scattering length, at unitarity we
have no scales, and so $g(\rho)\propto \rho^{5/3}$ is required on dimensional
grounds, reproducing the \gls{UFG} \gls{EoS}.  The normalization of the
time-evolution equation ensures Galilean covariance.

This modified \gls{GPE} corresponds to class of \glspl{DFT} known as
\glsreset{ETF}\gls{ETF} models.\footnote{The \gls{TF} approximation to a
  fermionic \gls{DFT} corresponds to applying the homogeneous \gls{EoS} at each
  point in space, introducing the external potential $V(x)$ as a spatially
  dependent chemical potential.  The ``extension'' here is corresponds to
  approximating the neglected kinetic term $\tau$ with gradients.  The kinetic
  term included in~\eqref{eq:ETF} along with the ``Weizsäcker'' term to be
  discussed below represent the lowest order expansion.  See~\cite{Brack:1997}
  for a discussion.}  In the absence of phase fluctuations, one can show that
the model~\eqref{eq:ETF} is equivalent to following local Hohenberg-Kohn
\gls{DFT}~\cite{Kim:2004a, *Kim:2004, Salasnich:2008b, *Salasnich:2008,
  *Salasnich:2008E}
\begin{gather}
  E[\rho] = \intd{3}{\vx}\left(
    \frac{\hbar^2}{32 m}\frac{(\vnabla\rho)^2}{\rho} 
    + V(\vx)\rho 
    + \xi\mathcal{E}_{FG}(\rho)\right),
\end{gather}
which is the more common form for the \gls{ETF} model.

The \gls{ETF} model reproduces the \gls{QHD} equations~\cite{Hu:2004} which
describe the evolution of the density and velocity fields $\rho$ and $\vv$:
\begin{subequations}
  \label{eq:QHD}
  \begin{gather}
    \begin{aligned}
      \Psi &= \frac{\rho}{\sqrt{2}} e^{2\I\phi}, &
      \vv &= \frac{\hbar\vnabla\phi}{m} 
      = \frac{\Psi^\dagger\I\vlrnabla\Psi}{2m\Psi^\dagger\Psi},
    \end{aligned}\nonumber\\[\baselineskip]
    \partial_t \rho + \vnabla\cdot(\rho \vv) = 0,\\
    -m \partial_t \vv = \vnabla\left(
      \frac{m v^2}{2} + V(\vx) + \xi E_F(\rho)
      - \frac{\hbar^2}{8m} \frac{\nabla^2 \sqrt{\rho}}{\sqrt{\rho}}
    \right).\label{eq:Bernoulli}
  \end{gather}
\end{subequations}
Note that~\eqref{eq:Bernoulli} -- the Bernoulli equation -- contains the
``quantum pressure'': singularities in this term are crucial for describing
quantum phenomena such as vortices.

There has been much discussion in the literature~\cite{Salasnich:2008b,
  *Salasnich:2008, *Salasnich:2008E, Zubarev:2009,
  Salasnich:2009,*Adhikari:2008a} about the addition of a Weizsäcker
term~\cite{Weizsacker:1935} to the functional in the form
\begin{gather}
  g(\rho) \rightarrow g(\rho) 
  - \frac{\hbar^2(1-4\lambda)}{32m}\frac{(\vnabla\rho)^2}{\rho}.
\end{gather}
This can be thought of as entering the equations of motions~\eqref{eq:E_UFG}
and~\eqref{eq:Bernoulli} through
\begin{gather}
  \xi E_F(\rho) \rightarrow \xi E_F(\rho)
  + \frac{\hbar^2(1 - 4\lambda)}{8m}\frac{\nabla^2\sqrt{\rho}}{\sqrt{\rho}}.
\end{gather}
There are two special values: $\lambda = 1/4$ corresponds to no Weizsäcker, and
$\lambda = 0$ cancels the quantum pressure term in the Bernoulli
equation~\eqref{eq:Bernoulli}, reducing the equations to classical hydrodynamics
of an irrotational and inviscid fluid. The absence of a Weizsäcker term gives
the expected analytic behavior of the condensate at the vortex
core~\cite{Forbes:2012a} and better describes the dynamics of colliding
superfluids~\cite{Ancilotto:2012, *Ancilotto:2012a}.  Hence in this paper we
will restrict to $\lambda = 1/4$.  The resulting \gls{ETF} model is described by
a single parameter -- the Bertsch parameter $\xi$.  To compare with the time
dependent \gls{SLDA} simulations of Ref.~\cite{Bulgac:2011b}, we will use their
value $\xi=0.42$.

The aim of this paper is to demonstrate the extent to which the \gls{ETF} can be
used to study the dynamics of the \gls{UFG} in place of the more computationally
expensive \gls{TDSLDA}.  Since the \gls{ETF} is tuned to match the \gls{UFG}
\gls{EoS}, it will by construction reproduce all related properties such as the
\gls{LO} (in energy and momentum) static and dynamic responses.  The non-trivial
validation comes when one considers higher orders and nonlinear effects.  We
consider three tests here: Sec.~\ref{sec:static-response}) the static response
at high wavevectors $q$; Sec.~\ref{sec:linear-response}) the dynamic linear
response at finite wavevector $q$ and frequency $\omega$;
Sec.~\ref{sec:nonlinear-response}) the nonlinear response by comparing with
\gls{TDSLDA} dynamics and experiments.

\begin{figure}[tbp]
  \includegraphics[width=\columnwidth]{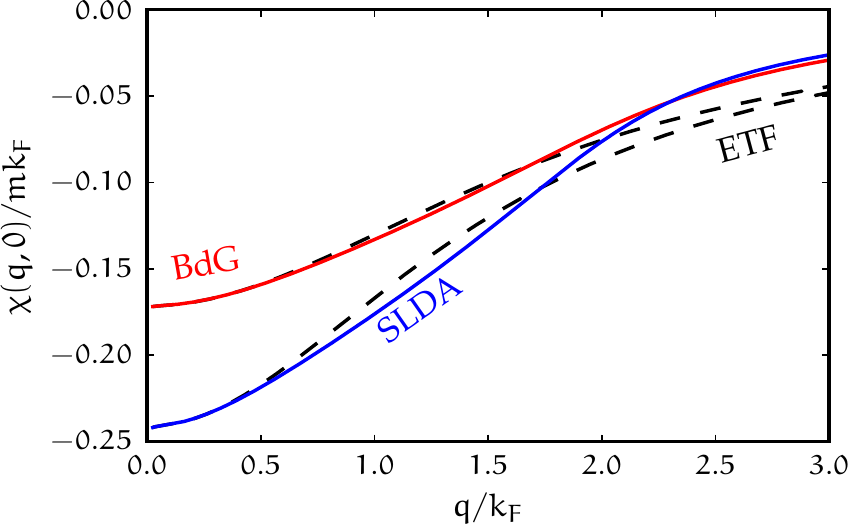}
  \caption{
    \label{fig:static_response}
    Static ($\omega=0$) response for two fermionic \glspl{DFT} and the
    corresponding \glspl{ETF}.  Upper curve: \gls{BdG} (upper solid red curve)
    with $\xi=0.5906\cdots$ and $\Delta=0.6864\cdots E_F$ (see
    also~\cite{Combescot:2006}).  Lower curve: \gls{SLDA} (lower solid blue
    curve) with $\alpha=1$, $\xi=0.42$, and $\Delta = 0.502E_F$ (to
    match~\cite{Bulgac:2011b}). The \glspl{ETF} (dashed black curve) have their
    single parameter $\xi$ tuned to match the respective fermionic theories, and
    consequently match at $q=0$ where the response (the compressibility) is
    determined by the \gls{EoS}. The curvature for small $q$ is incidentally
    numerically very similar for the corresponding theories (see
    Section~\ref{sec:linear-response}).  The deviations for larger $q$ give an
    estimate of how well the \glspl{ETF} can model the fermionic theories.}
\end{figure}

\section{Static Response}\label{sec:static-response}\noindent
The static \gls{ETF} model has been compared with \gls{QMC} results for the
harmonically trapped \gls{UFG}~\cite{Salasnich:2008b, *Salasnich:2008,
  *Salasnich:2008E}.  Comparison with recent \gls{QMC}
results~\cite{Forbes:2012a} demonstrate that it exhibits the correct qualitative
asymptotic behaviour in the thermodynamic limit, reproducing the asymptotic form
predicted by the low-energy superfluid \gls{EFT}~\cite{SW:2006}, but fails for
small systems. This failure is expected since the \gls{ETF} lacks the kinetic
density $\tau$ required to reproduce the fermionic shell structure resulting
from the Pauli exclusion principle.

The linearized static density response $\chi_{\rho}(q, \omega=0)$ for wavevector
$q$ is defined by considering how the density changes in response to a small
cosine modulation:
\begin{align*}
  V_{R}(x) &= \delta \cos(qx), &
  \rho(x) &= \rho_0 + \chi_\rho(q, \omega=0)\; \delta\cos(qx).
\end{align*}
The static response of the \gls{ETF} and \gls{SLDA} are compared in
Fig.~\ref{fig:static_response}.  The \gls{EoS} fixes the value
$\chi_\rho(q\rightarrow 0, \omega=0) = \partial n/\partial \mu$, but the
\gls{ETF} matches that of the fermionic \gls{SLDA} quite well, even for large
wave-vectors. Although we do
not consider such corrections here, it should be possible to add gradient
corrections to the \gls{ETF} to improve this agreement (being careful not to
affect the analytic structure of vortices etc.).

\begin{figure}[tbp]
  \includegraphics[width=\columnwidth]{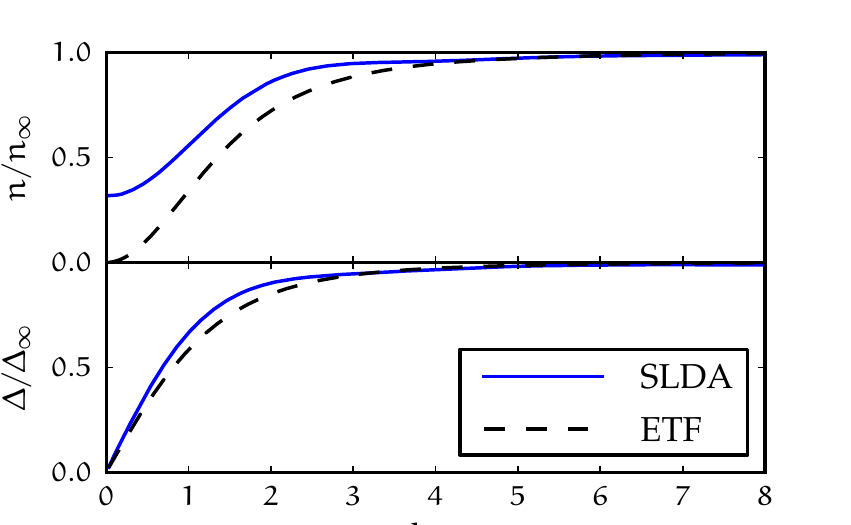}
  \caption{\label{fig:vortex_profile} Structure of a single static vortex in the
    \gls{SLDA}~\cite{BY:2003} (solid blue curve), and in the matching \gls{ETF}
    (dashed black curve).  We compare only with parameter set II
    from~\cite{BY:2003} which has unit inverse effective mass $\alpha = m/m_* =
    1$ and parameters tuned so that $\xi=0.44$ while the energy of the normal
    state is $\xi_N=0.54$ (this gives a somewhat low pairing gap $\Delta \approx
    0.3718 E_F$).  We do not consider the $\alpha\neq 1$ vortex for parameter
    set I in~\cite{BY:2003} which is missing the corrections that restore
    Galilean invariance~\cite{Bulgac:2011}.}
\end{figure}

To compare the static response in the nonlinear regime, we consider the
structure of a single vortex in Fig.~\ref{fig:vortex_profile}. This demonstrates
one major limitations of the \gls{ETF} model which imposes an artificial
relationship between the square of the order parameter and the density $\rho =
2\abs{\Psi}^2$. In the fermionic theory, the relationship between $\Delta$ and
$\rho$ are determined as independent sums of the single-particle wavefunctions:
the relation, $\rho = 2\abs{\Psi}^2$ only becomes valid for fermions in the deep
\gls{BEC} regime.  In the \gls{UFG}, the vortex cores have a non-zero density
(often thought of as ``normal'' fermionic modes occupying the vortex core where
the superfluid condensate vanishes), while the \gls{ETF} by construction has
zero-density wherever the condensate $\Psi=0$ vanishes.

% This full-width figure needs to be included in text from the page *before* it
% should appear in the document.
\begin{figure*}
  \includegraphics[width=\textwidth]{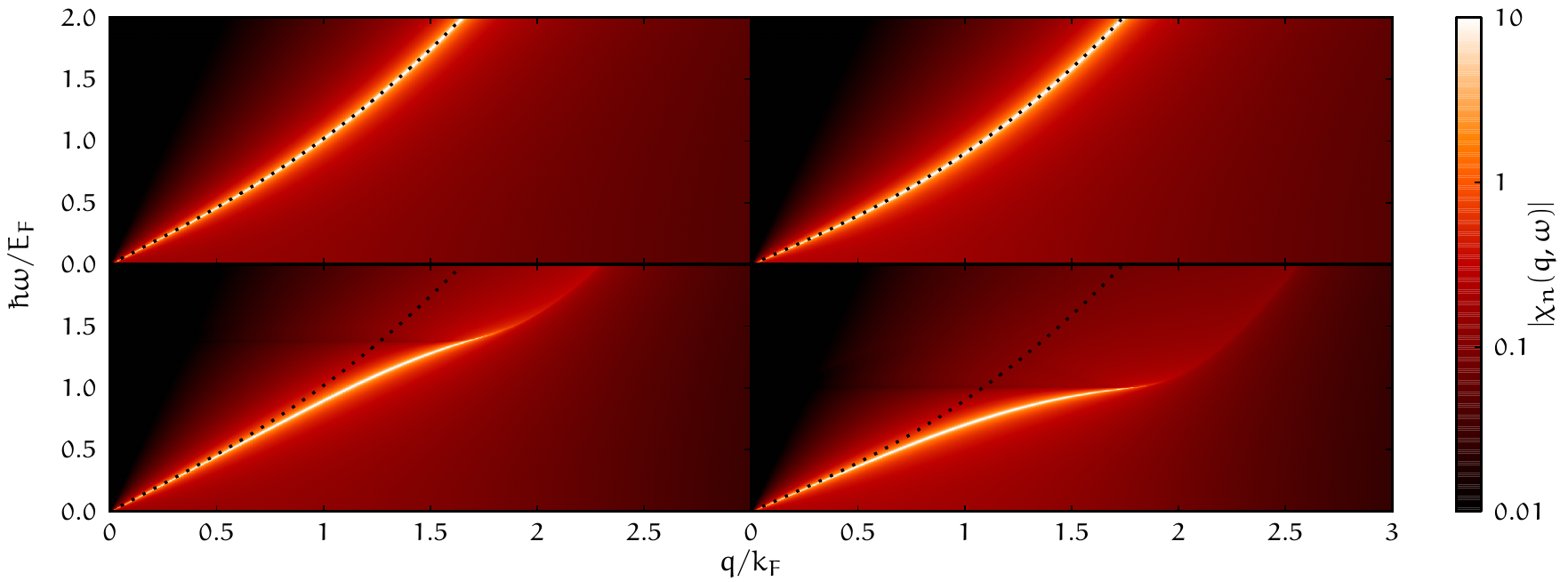}
  \caption{
    \label{fig:linear_response}
    Comparison of the linear response for the \gls{ETF} (top) and two fermionic
    \glspl{DFT} (bottom).  The linear response of the \gls{BdG} which has
    $\xi=0.5906\cdots$ and $\Delta=0.6864\cdots E_F$ (see
    also~\cite{Combescot:2006}) is on the lower left; the linear response of
    the \gls{SLDA} tuned to $\xi=0.42$ and $\Delta = 0.502E_F$ to
    match~\cite{Bulgac:2011b} is on the lower right.  The \gls{ETF} has only
    the single tunable parameter $\xi$, which is chosen to match the
    corresponding fermionic theory in the panel immediately below. \P~The
    bosonic \gls{ETF} reproduces the low-frequency response, but breaks down
    for $\omega \approx 2\Delta$ at the pair-breaking threshold. The slope of
    the phonon dispersion relationship is reproduced near the origin, but 
    the curvature differs between the fermionic and bosonic
    theories.
  }
\end{figure*}

This core occupation also appears in solitons, giving rise to a change in the
oscillation period for solitons in a quasi-1\mysc{d} harmonic trap from
$T\approx\sqrt{2}T_z$~\cite{%
  Busch:1998, % Original calculation
  *Busch:2000, % First published version by original authors.
  Muryshev:1999a, % Appeared almost at the same time
  Fedichev:1999, % Discusses dissipation
  Konotop:2004, % Grey solitons
  Becker:2008, *Weller:2008} % Experiments
in the bosonic systems (reproduced by the \gls{ETF} model) to
$T\approx\sqrt{3}T_z$ in the fermionic \glspl{DFT} (\gls{BdG}~\cite{Scott:2011}
and \gls{SLDA}~\cite{Scott:2011pc}).  Thus, bosonic and fermionic simulations
are qualitatively, but not quantitatively, similar when describing these
types of dynamics.  Note that recent experiment~\cite{Yefsah:2013}
suggest that solitons in the \gls{UFG} might have a significantly longer period
$T\approx 10T_z$, but this has been resolved by identifying the observations
with vortex rings~\cite{Bulgac:2013d}.

Related to the deficiency in properly describing the core density, we note that
unitary evolution of the \gls{ETF} implies that
\begin{gather}\label{eq:gap_conservation}
  \pdiff{}{t}\intd{3}{\vx}\Psi^\dagger(\vx,t) \Psi(\vx,t) = 0.
\end{gather}
This means that, not only is the total particle number conserved (which is
physical), but the integrated ``gap'' is also conserved
$\smash{\partial_t\intd{3}{\vx}\abs{\Delta}^2 = 0}$.  In fermionic systems,
pair-breaking excitations will reduce the gap, resulting in a mixture of
superfluid and normal fluid; in highly excited systems the superfluid may vanish
completely.  The \gls{ETF} on the other hand does not admit this behaviour, and
even highly excited systems will still have a rapidly fluctuating but non-zero
order parameter.  The degree to which the integrated gap is conserved during the
evolution of a fermionic system provides a useful measure of how successfully
the \gls{ETF} can model the corresponding evolution.  (We shall explore this
further in Fig.~\ref{fig:DeltaConservation}.)

Despite the fact that the resulting \gls{ETF} contains only a single parameter
(compared with the three independent parameters of the \gls{SLDA}), it still
qualitatively reproduces many response properties.  This qualitative agreement
is a somewhat fortuitous consequence of the best-fit parameter values.  From the
point-of-view of the \gls{ETF}, the \gls{UFG} contains two independent length
scales: the inter-particle spacing set by the density, and the coherence length
set by the gap. This is demonstrated by the failure of the \gls{ETF} to capture
the core structure of a vortex.  Thus, while the present concordance of the
\gls{SLDA} and \gls{ETF} is fortuitous, it may turn out that the \gls{SLDA}
requires further gradient corrections~\cite{Forbes:2012a} (a result that is
still awaiting further \textit{ab initio} confirmation).  If this correction
turns out to be significant, then one might have to introduce gradient
corrections in a more complicated form (compared to the simple Weizsäcker term)
that does not spoil vortex structure and collision dynamics.  Such corrections
will be non-universal (i.e. must have a different form for small densities than
for large densities) and probably most conveniently accounted for in two-fluid
model with an additional ``normal'' component that can populate the vortex
core.  The approximation to the \gls{BdG} discussed in~\cite{Simonucci:2013} may
shed some light on the nature of these types of corrections.

\section{Linear Response}\label{sec:linear-response}\noindent
We now consider dynamical systems.  For small fluctuations one can simply
compare the linear response of the \gls{ETF} with that of the Fermi systems.
We compute the response of the system to an external time-dependent
perturbation in the limit of small $\delta$:
\begin{gather*}
  V_{R}(x,t) = \delta \Re\bigl[ e^{\I(qx + \omega t)}\bigr], \\
  \rho_{R}(x, t) = \rho_0 + \delta\Re\bigl[\chi_{n}e^{\I(qx + \omega t)}\bigr]
  + \order(\delta^2).
\end{gather*}
The magnitude of the resulting response $\abs{\chi_n}$ is shown in
Fig.~\ref{fig:linear_response} for the \gls{BdG} and
\gls{SLDA} and compared with the response for the corresponding \gls{ETF} model
tuned to match the value of $\xi$.

The response at low frequencies is dominated by the pole associated with the
superfluid phonon.  This may be computed analytically for homogeneous matter in
the \gls{ETF}:
\begin{gather}
  \omega_{\text{phonon}} = \sqrt{
    \left(\frac{\hbar q^2}{4 m}\right)^2 
    + \frac{2 q^2}{3 m} \xi E_F
  }
  = c_s q + \order(q^3),
\end{gather}
where $c_s = \sqrt{\xi/3} v_F$ is the sound speed and $v_F = \hbar k_F/m$ is the
Fermi velocity. At small momenta, the $f-$sum rule~\cite{GV:2005} ensures that
the residue of the pole in the bosonic and fermionic theories at low $q$ is
equal to $-\pi \rho_0 q^2\hbar^2/(2m\omega)$.

The low-energy properties of these theories can be characterized by a superfluid
\gls{EFT} for the \gls{UFG}~\cite{SW:2006} (also see~\cite{Manes:2009,
  *Schakel:2011}). At \gls{LO}, the theory is characterized by the Bertsch
  parameter $\xi$ which determines the equation of state.  Two new coefficients appear at \gls{NLO},\footnote{The
  coefficients $\cSR$ and $c_\omega$ are ``natural'' in the sense that $\cSR
  \approx c_\omega \approx 1$. Different notations are used in~\cite{SW:2006}
  and ~\cite{Manes:2009}: both use $\xi =2^{5/3}/(15 c_0 \pi^2)^{2/3}$:
  \begin{align*}
    \frac{\cSR}{-6\pi^2(2\xi)^{3/2}} &\text{ is } 
    2c_1 - 9c_2 \text{ in Ref.~\cite{SW:2006}, but } 
    2c_1 \text{ in Ref.~\cite{Manes:2009},}\\
    \frac{c_\omega}{-6\pi^2(2\xi)^{3/2}} &\text{ is } 
    2c_1 + 3c_2 \text{ in Ref.~\cite{SW:2006}, but }
    2c_1 - 6c_2 \text{ in Ref.~\cite{Manes:2009}.}
  \end{align*}} which we shall denote $\cSR$ and $c_\omega$
following~\cite{Forbes:2012a}, that characterize the low-energy static and
dynamic properties respectively. These coefficients characterize the phonon
dispersion $\omega_q$ and static response $\chi(q, \omega=0)$:
\begin{gather}
  \label{eq:SonWingatePhonon}
  \omega_q = c_s q\left[
    1 + \frac{c_\omega}{24\xi}\frac{q^2}{k_F^2}
    + \order(q^4\ln{q})\right],\\
  \label{eq:SonWingateStaticResponse}
  \chi(q, \omega=0) = \frac{-mk_F}{\hbar^2\pi^2\xi}\left[
    1 - \frac{\cSR}{12\xi}\frac{q^2}{k_F^2} + \order(q^4\ln{q})
  \right]
\end{gather}
Matching with the linear response of the \gls{ETF} gives $c_\omega = \cSR =
9/4$.  This is qualitatively consistent with estimate of these parameters from
the $\epsilon$-expansion~\cite{Rupak:2008fk} (expanding in spatial dimension:
$\epsilon = 4 - d$) which finds $\cSR \approx 8/5 + \order(\epsilon^2)$ and
$\cSR \approx c_\omega + \order(\epsilon^2)$.\footnote{To compare
  with~\cite{Rupak:2008fk}, note that their $c_s\equiv \cSR/2\xi$.}
The \gls{BdG} mean-field theory~\cite{Manes:2009} finds quite different values,
$\cSR = 7/3$ and $c_\omega = 0.7539$. Interestingly, for $\alpha=1$
the \gls{SLDA} gives $\cSR=7/3$ independent of the values of $\beta$ and
$\gamma$ in Eq~\ref{eq:DF_SLDA} (or equivalently $\eta$ and $\xi$). The value of
$c_\omega$ in \gls{SLDA} is not quite as robust. For fixed $\eta=0.502$ and
$\alpha=1$, $c_\omega$ changes from $-0.255$ to $0.055$ as $\xi$ is reduced from
$0.42$ to $0.37$ (i.e.\@ from the value used in Fig.~\ref{fig:linear_response}
to the current best fit value). The value for $\cSR \approx \num{1.5(3)}$
follows from an analysis of gradient corrections to harmonically trapped
gases~\cite{Forbes:2012a}.

The value of $c_\omega$ determines the curvature of the phonon dispersion. As is
clear from Fig.~\ref{fig:linear_response} the \gls{ETF} gives a large positive
curvature for the dispersion. In contrast the dispersion curves for \gls{BdG}
and \gls{SLDA} appear relatively linear and eventually curve downward. This is a
combination of two effects. First, $c_\omega$ is smaller in the fermionic
theories (negative for \gls{SLDA}). Second, higher-order effects in $q/k_F$ pull
the curves down as one approaches the pair-breaking threshold. For the \gls{BdG}
this implies the existence of a point of inflection at $q/k_F \approx
\num{0.53}$.

That the \gls{ETF} has no transverse response, which can be expressed in terms
of the difference $\cSR - c_\omega$, represents another shortcoming of the
model.  As argued in~\cite{SW:2006}, the transverse response should be positive
to ensure stability with respect to a spontaneous generation of currents or
formation of inhomogeneous condensates.

To end this section, we consider the numerical values using
$\xi=\num{0.374}$~\cite{Forbes:2012}:
\begin{align*}
  \omega_q &\propto 1 + 0.11 c_\omega \frac{q^2}{k_F^2}, &
  \chi(q) &\propto 1 - 0.22 \cSR\frac{q^2}{k_F^2}.
\end{align*}
Since $\cSR \approx c_\omega \sim 1$, we see that the prefactor multiplying the
correction to the leading order dynamics is somewhat small. Thus the low-energy
dynamics are rather insensitive to the limitation that $\cSR - c_\omega$
vanishes in the \gls{ETF} and that $c_\omega$ is somewhat larger than in
fermionic theories.  The partly explains the success that the \gls{ETF} enjoys
at low-energy. 

\begin{figure*}[tb]
  \includegraphics[width=0.98\textwidth]{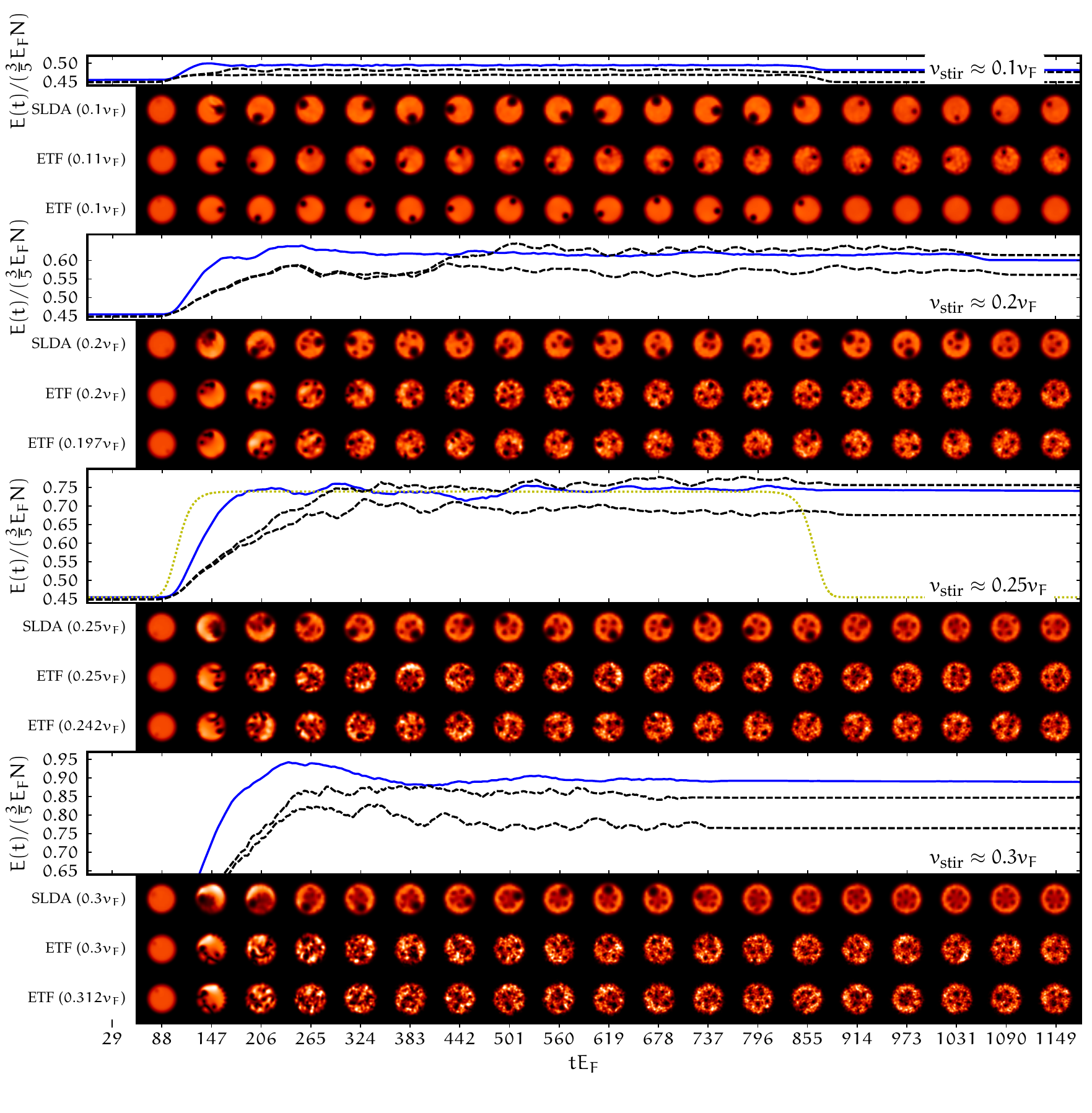}
  \caption{\label{fig:E_vs_t}(color online) %
    Stirring simulations. The curves show the energy per particle as a function
    of time for various simulations for increasing stirring speeds:
    $v_{\text{stir}}\approx 0.1v_F$ (top), $v_{\text{stir}}\approx0.2v_F$,
    $v_{\text{stir}}\approx0.25v_F$ (middle), and $v_{\text{stir}}\approx0.3v_F$
    (bottom).  Sample density profiles are shown below the $x$-axis starting with the
    \gls{SLDA} simulation from~\cite{Bulgac:2011b}, followed by the \gls{ETF}
    simulation(s).  The plots each have two \gls{ETF} simulations, one with
    exactly the same $v_{\text{stir}}$ as the \gls{SLDA}, and another with a
    slightly different $v_{\text{stir}}$ that produces the same total number of
    vortices and give the energy curves (dashed, black online) immediately
    below the \gls{SLDA} energy curves (solid, blue online).  The \gls{SLDA}
    simulations use a $32^2$ lattice while the \gls{ETF} simulations use a
    $64^2$ lattice. The light dotted (yellow) curve in the third plot shows the
    strength of the stirring potential (in arbitrary units) as it is turned on,
    held, then turned off.}
\end{figure*}

\section{Nonlinear Response}\label{sec:nonlinear-response}% (keep %)
From the previous analysis, we expect the \gls{ETF} to provide a reasonable
description of small-amplitude fermionic dynamics as long as one does not push
the system to the pair-breaking threshold: i.e.\@ for slowly varying external
potentials. In the nonlinear regime, the disagreement grows as evolution in the
\gls{ETF} transfers energy from the small momentum modes to the higher momentum
modes. The high-momentum response is dominated by the phonon dispersion, so this
tends to create excitations in the pair-breaking regime where the \gls{ETF}
breaks down.

The transfer of energy to higher momenta results from the nonlinear interaction
term which acts as a wave-vector multiplier. The result is that \gls{ETF} simulations tend
to be more noisy than the corresponding \gls{TDSLDA} simulations. To see this in
a concrete example, we directly compare the dynamics of the \gls{ETF} with that
of the \gls{TDSLDA} using the same trapping potential and time-dependent
stirring potential as in~\cite{Bulgac:2011b}.

The setup is as follows: The cloud is prepared in the ground state of a
two-dimensional axially symmetric flat-bottomed trap of radius $R$ (precise
details of the potential etc.\@ are given in Appendix~\ref{sec:trap-params}). A
repulsive potential at a distance $r_{\text{stir}}$ from the centre rotates with
constant angular frequency $\omega_{\text{stir}} =
v_{\text{stir}}/r_{\text{stir}}$.  This is gradually turned on, left on for
$n_{\text{stir}}\approx 10$ rotations, then gradually turned off. These
simulations are quasi--two-dimensional and have translational symmetry along the
trap axis: the fermionic simulation discretize the wavefunctions on a $32^2$
two-dimensional lattice.  It will turn out that simulating the \gls{ETF} on
larger ($64^2$) lattices better reproduces features of the fermionic theory --
most likely due to the transfer of energy to higher-momentum modes discussed
above. Figure~\ref{fig:E_vs_t} summarizes some sample results.

\begin{table}[tb]
  \caption{\label{tab:vortices_vs_vAurel}%
    Number of vortices created after ten revolutions of a stirring potential as
    a function of the stirring velocity $v_{\text{stir}}/v_F$ where $v_F$ is the
    Fermi velocity at the centre of the trap.  The second column (\gls{SLDA})
    shows the results of~\cite{Bulgac:2011b} on a $32^2$ lattice, while the
    third and fourth columns show the corresponding results for the \gls{ETF}
    with lattices of $32^2$ and $64^2$ points respectively. For the higher
    velocities, the $32^2$ \gls{ETF} simulations are too noisy to admit
    an accurate count of the vortices.
  }
  \begin{ruledtabular}
    \begin{tabular}{dddd}
      \multicolumn{1}{c}{$v_{\text{stir}}/v_F$} & 
      \multicolumn{1}{c}{\gls{SLDA} ($32^2$)} & 
      \multicolumn{1}{c}{\gls{ETF} ($64^2$)} & 
      \multicolumn{1}{c}{\gls{ETF} ($32^2$)}
      \\ 
      \hline
      0.1   &  1 & 0 & 0\\
      0.11  &  - & 1 & 1\\
      0.197 &  - & 3 & 2\\
      0.2   &  3 & 4 & 3\\
      0.242 &  - & 5 & 2\\
      0.25  &  5 & 6 & 2\\
      0.3   &  6 & 5 & \multicolumn{1}{c}{(noise)}\\
      0.312 &  - & 6 & \multicolumn{1}{c}{(noise)}\\
      0.35  &  7 & 7 & \multicolumn{1}{c}{(noise)}\\
      0.40  &  9 & 9 & \multicolumn{1}{c}{(noise)}\\
    \end{tabular}
  \end{ruledtabular}
\end{table}

In Table~\ref{tab:vortices_vs_vAurel} we compare the number of vortices
generated after four (for $v_{\rm{stir}}=0.1$, $0.11$) or ten revolutions (for
all others) of the stirring potential, and demonstrates the qualitative
agreement between the \gls{ETF} model and the \gls{TDSLDA}. For small
velocities, there are some minor disagreements: at $v_{\text{stir}} = 0.1$, the
\gls{ETF} does not produce a vortex, but does produce one for $v_{\text{stir}} =
0.11$.  This $\approx 10\%$ difference might be due to differences in the static
response of the \gls{ETF} and the \gls{SLDA}.  For example, the pinning
potential creates a larger depletion in the \gls{SLDA} (visible in
Fig.~\ref{fig:E_vs_t}), thereby exciting regions closer to the edge of the trap
where the density is lower.  These regions have a slower critical velocity,
allowing a vortex to be nucleated more easily than in the \gls{ETF}.  One might
consider tuning the potential or model as suggested in~\cite{Bulgac:2013a} to
study vortex-pinning interactions to match the density depletion in the
\gls{ETF} to that in the \gls{SLDA}, but have not performed any such tuning
here.%
\exclude{This might have been anticipated from the static response in
  Fig.~\ref{fig:static_response} which is suppressed in the \gls{ETF} for
  $q\gtrapprox 2k_F$ (the pinning potential has $k_Fr_{\text{pin}} \sim 2$), }%
\exclude{Removed because the pinning site only has support up to $q\lesssim 1.5
  k_F$ I think the extent of the depletion here is related to nonlinear effects
  which also explains why the effects are different for different $v_stirs$.}

\begin{figure}[htb]
  \includegraphics[width=\columnwidth]{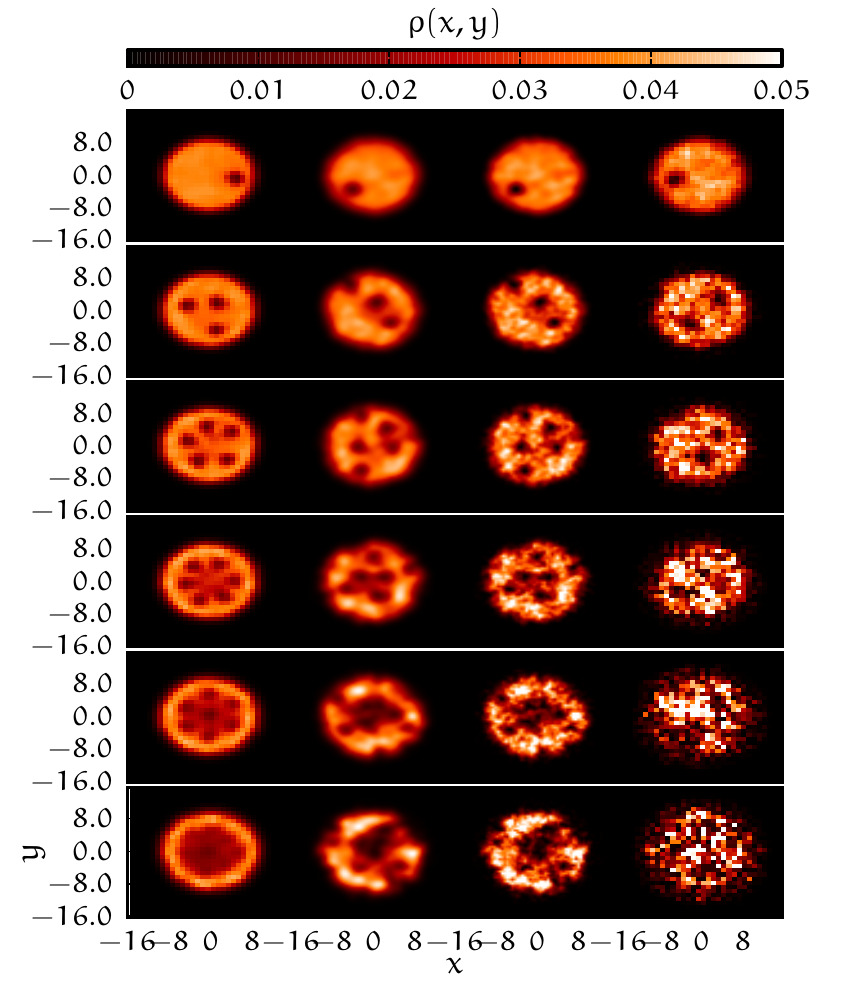}
  \caption{\label{fig:compare_resolutions}(Color online) %
    Comparison of final state densities (moving from left to right) for $N=32$
    \gls{TDSLDA}, a smoothed version of $N=64$ \gls{ETF}, the raw result for
    $N=64$~\gls{ETF} and the raw result for the $N=32$ \gls{ETF}. The
    simulations are for $v_{\rm{stir}}\approx 0.1,\, 0.2,\, 0.3,\, 0.35,\,
    0.40$ (moving from top to bottom), taking the $v_{\rm{stir}}$ for the
    \gls{ETF} from Table~\ref{tab:vortices_vs_vAurel} which gives the same
    number of vortices as the \gls{SLDA}. The \gls{TDSLDA} vortices are arranged
    in a regular pattern in the final state.  The $N=32$ \gls{ETF} is too
    noisy. The higher resolution ($N=64$)~\gls{ETF} are qualitatively more
    similar, especially after smoothing, but the vortices are not as regularly
    arranged as the \gls{TDSLDA} simulations. The density was smoothed by
    convolving with a two-dimensional Gaussian smearing function of spatial
    width $0.75/k_F$.}
\end{figure}

As one increases the rotation rate, one finds that the $32^2$ simulations depart
significantly -- these essentially develop short-wavelength noise due to the
aforementioned amplification of short-wavelength modes to a point where
identifying vortices becomes impossible.  The problem here is essentially that
significant phonon ``noise'' coexists on the same length-scale as the vortex
core.  Increasing the resolution resolves this issue by providing a separation
of scales between the phonon ``noise'' and the larger structure of the vortices.
A comparison of the final states obtained for various resolutions is shown in
Fig.~\ref{fig:compare_resolutions}.

\exclude{
\begin{figure}[tb]
  \includegraphics[width=\columnwidth]{spectra_evolution_density}
  \caption{\label{fig:spectra_evolution}(color online) %
    Comparing the evolution of power spectra in the \gls{ETF} for three
    different stirring velocities. The lighter color (yellow online) corresponds
    to the smaller wave number ($q/k_F=1$) and the darker color (green online)
    is for twice that. From the lower panels, we see that the amplitude of the
    ($q/k_F=2$) rises more gradually compared to the longer wavelength
    ($q/k_F=1$) mode, but eventually saturates at a value comparable to it
    (especially for the faster stirring speeds). The upper panels verify that
    smaller wavelengths have power at higher frequency.}
\end{figure}
}

The rough final agreement in vortex number between the two theories follows
mainly from the superfluidity of the system.  In order to support a rotational
current with a fixed stirring velocity $v_{\text{stir}}$ at the specified
radius, the system must carry enough angular momentum, hence they must have
(roughly) a certain number of vortices.  Once the system achieves a rotational
flow with $v = v_{\text{stir}}$ an equilibrium is established and no further
energy is transferred from the stirrer to the system.  From
Fig.~\ref{fig:E_vs_t} we see that the overall energy scale for a given number of
vortices is roughly equal in the \gls{ETF} and the \gls{SLDA} (the final
\gls{ETF} energies are systematically slightly smaller than the final \gls{SLDA}
energies for the same number of vortices). This is because the equation of
state of the two systems are the same and energies are the kind of bulk
property for which the \gls{ETF} can be trusted.

From the details in Fig.~\ref{fig:E_vs_t}, this bulk agreement is apparent.  In
addition, one sees detailed qualitative agreement in the dynamics for lightly
excited systems. For example, the single vortex produced for $v_{\text{stir}}
\approx 0.1v_F$ behaves almost identically in both the \gls{TDSLDA} and
\gls{ETF}. While the stirrer is ``on'', the vortex is closely attached to
it. When the stirrer has ``switched off'', the vortex continues (roughly)
rotating around the center of the trap with an angular velocity
$\omega\sim{\hbar}/{(2m(R^2-r_{\text{stir}}^2))}$ determined by the background
superfluid velocity at the vortex induced by the trap~\cite{PS:2002}.

For higher velocities, however, many qualitative differences between the
\gls{TDSLDA} and \gls{ETF} dynamics become apparent. Most obviously, the energy
transfer to the \gls{ETF} is significantly slower in the \gls{ETF} than in the
\gls{TDSLDA}.  Another obvious feature is that the \gls{ETF} vortex lattice
does not ``crystallize'' as it does in the \gls{TDSLDA}.  This is similar
behaviour to the \gls{GPE} where crystallization is known to require the
addition of dissipative mechanisms as in the \gls{SGPE}
(see~\cite{Gardiner:2002} and references therein). 

\begin{figure}[tb]
  \includegraphics[width=\columnwidth]{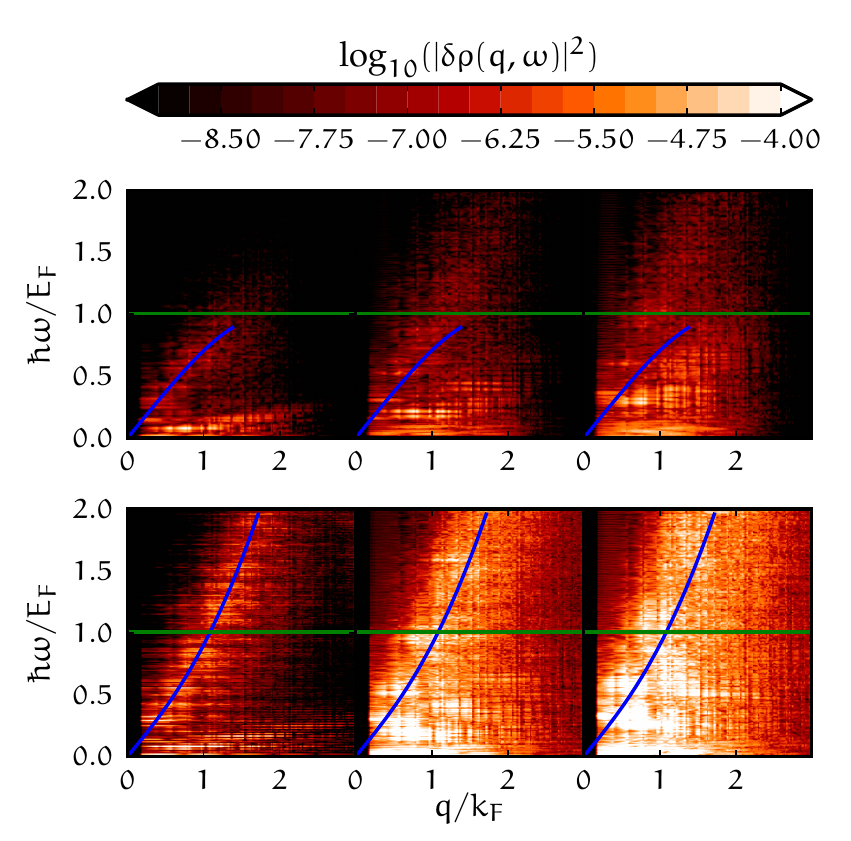}
  \caption{\label{fig:PowerSpectrums}(color online) % 
    We compare the power spectra of the fluctuations of density in the bosonic
    simulations (lower panel) and the fermionic simulations (upper panel). For
    the bosonic (fermionic) simulations we consider stirring velocities
    $v_{\text{stir}}=0.11 v_F$ ($0.10 v_F$), $v_{\text{stir}}=0.197 v_F$ ($0.20
    v_F$), and $v_{\text{stir}}=0.242 v_F$ ($0.25 v_F$), going from left to
    right. The spatial Fourier transform is taken over entire simulation volume
    and the temporal transform is taken over the time after the stirring
    potential is turned off. The solid horizontal line (green online)
    corresponds to the pair breaking threshold, $\omega/E_F=2\eta$. For the
    \gls{SLDA} simulations, there is little strength above the pair breaking
    threshold. For the \gls{ETF} simulations with the smallest velocity, most
    of the power is concentrated in the low frequencies. For higher velocities,
    there is significant power near and above the pair-breaking threshold. The
    curves (blue online) correspond to the phonon dispersion relation.
  }
\end{figure}

\begin{figure}[tb]
  \includegraphics[width=\columnwidth]{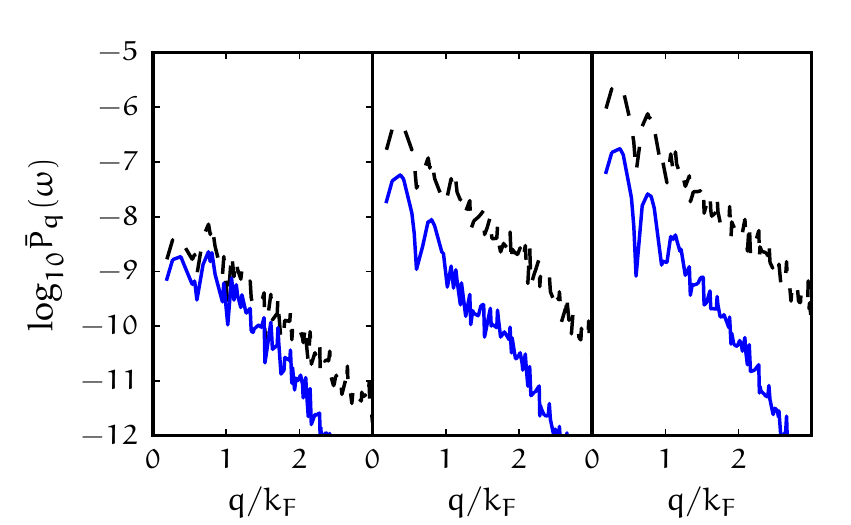}
  \caption{\label{fig:spatial_spectra}(color online) %
     Comparison of the power spectrum (averaged over the time after ``switch
     off'') for different momentum modes for the \gls{TDSLDA} and the
     \gls{ETF}.  The fluctuations in the \gls{SLDA} drop down at a faster rate
     than the \gls{GPE}.}
\end{figure}

The nonlinear nature of the \gls{ETF} implies that even if initially only
long-wavelength modes are excited (for example, in our simulations the stirring
potential only only has support for momenta up to $q/k_F\lesssim 1.5$), energy
can be transferred to short-wavelength modes. This phenomenon is common to a
variety of nonlinear non-dissipative systems, for instance optical systems,
cold plasmas, and bosonic superfluids described by the
\gls{GPE}~\cite{Wan:2007}.  The difficulty this presents with the \gls{ETF} is
that the phonon pole continues to extend to both high momenta and high
frequencies, whereas in the \gls{SLDA}, the pole is replaced by a branch cut at
the pair-breaking threshold (see Fig.~\ref{fig:linear_response}).  Thus, while
large wavelength modes seem to decay in the \gls{SLDA}, they persist (see
Fig.~\ref{fig:spatial_spectra}) in the \gls{ETF} giving rise to noisy
simulations that cannot reproduce features such as the relaxation of vortex
lattices (see Fig.~\ref{fig:compare_resolutions}).

To contrast the situation from the \gls{SLDA} we compare the power-spectra of
the density perturbations in Fig.~\ref{fig:PowerSpectrums}.  These spectra are
computed after the stirring potential is turned off and demonstrate that the
majority of the power lies along the phonon dispersion.  These simulations also
have vortices, which add power at low frequencies (one can think of a vortex as
a collection of virtual photons).  Note that in the \gls{ETF}, even the slowest
simulation $v_{\text{stir}} = 0.11v_F$ has energy above the pair-breaking
excitation, demonstrating the amplification of short-wavelength modes.

All of this evidence is commensurate with the fundamental failure of the
\gls{ETF} to properly describe pair-breaking excitations above $\omega >
2\Delta$ that appear to be present in all simulations (except the vortex-less
$v_{\text{stir}} = 0.1$ simulation).  In the \gls{TDSLDA}, these excitations
break superfluid pairs, transferring energy to the normal component of the fluid
which is absent in the \gls{ETF}.  This provides a damping mechanism for the
superfluid in the \gls{TDSLDA} that allows the vortex lattice to crystallize.
In the \gls{ETF}, these excitations must remain in the superfluid and scatter
off of the vortices, preventing the lattice from crystallizing.  

\begin{figure}[tb]
\includegraphics[width=\columnwidth]{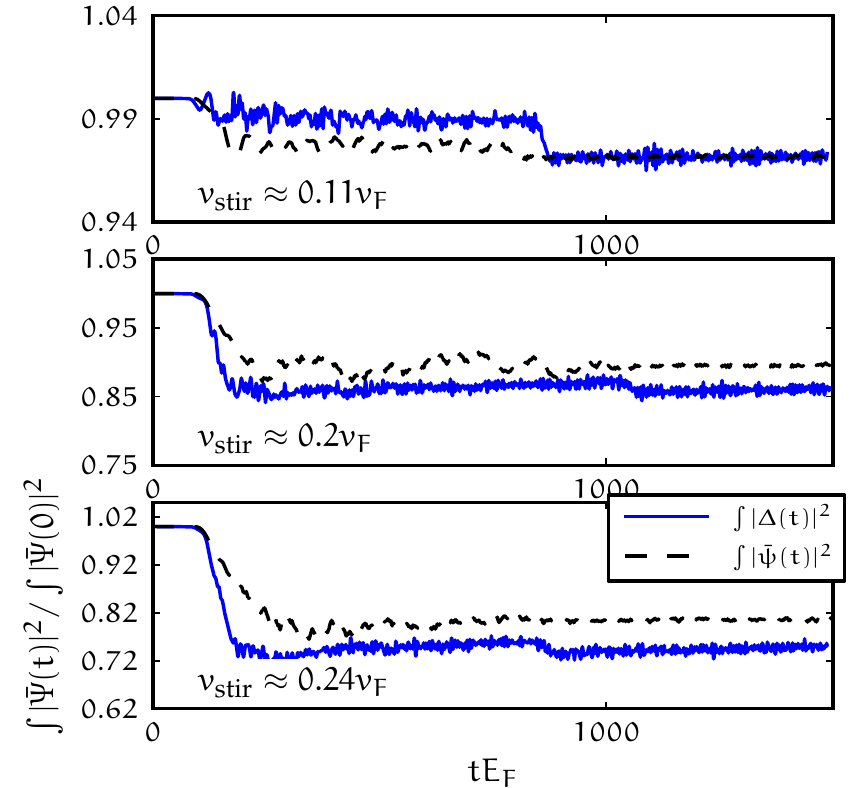}
\caption{\label{fig:DeltaConservation}(color online) % 
  Conservation of the integrated squared pairing gap (squared smoothed $\psi$)
  for the simulations for $v_{\text{stir}}=0.1 v_F$, ($v_{\text{stir}}=0.11
  v_F$) $v_{\text{stir}}=0.2 v_F$, ($v_{\text{stir}}=0.197 v_F$), and
  $v_{\text{stir}}=0.25 v_F$ ($v_{\text{stir}}=0.242 v_F$) for \gls{TDSLDA}
  (\gls{ETF}). The wavefunction was smoothed by convolving with a
  two-dimensional Gaussian smearing function of spatial width $0.75/k_F$. }
\end{figure}

To check this, we can consider the superfluid order parameter $\Delta$.
Pair-breaking effects reduce the amount of superfluid, resulting in a decrease
in the total integrated gap as the \gls{TDSLDA}, whereas the corresponding
quantity in the \gls{ETF}, the particle number, remains
conserved~\eqref{eq:gap_conservation}:
\begin{align}\label{eq:integrated_gap}
  \text{\gls*{TDSLDA}: }&\int \d^{3}{\vect{x}}\; \abs{\Delta}^2 & &\text{vs.} &
  \text{\gls*{ETF}: }&\int \d^{3}{\vect{x}}\; \abs{\Psi}^2.
\end{align}
To realize pair-breaking physics in an \gls{ETF}-like model, one needs to
introduce an additional thermal ``normal'' component to the system, transferring
energy and mass to this as excitations exceed the pair-breaking threshold.  To
test the validity of this notion, we compare in Fig.~\ref{fig:DeltaConservation}
the evolution of the integrated pairing gap~\eqref{eq:integrated_gap} in the
\gls{TDSLDA} with the integrated order parameter in the \gls{ETF} after
coarse-graining the field $\Psi$ with a filter that removes excitations above $q
\gtrsim 1.3k_F$.  (We simply smoothed the $64^2$ simulation with a two-dimensional Gaussian smearing function of spatial
width $0.75/k_F$.)

The qualitative agreement here shows that this characterization of the
superfluid to normal conversion is reasonable. This is visually confirmed in
Fig.~\ref{fig:compare_resolutions} where we also include a coarse-grained
representation of the density (smoothing now the density $\rho = 2\abs{\Psi}^2$
rather than $\Psi$). 

A similar coarse graining of the evolved \gls{ETF} was performed
in~\cite{Ancilotto:2012, *Ancilotto:2012a} to compare with the shock-wave
experiment~\cite{Joseph:2011}.  The agreement there confirms this picture that
the \gls{ETF} is suitable for modelling bulk dynamical properties.  Note,
however, that the difference in dynamics here is in contrast with the implied
claim of Refs.~\cite{Ancilotto:2012, *Ancilotto:2012a} that the coarse graining
is simply needed to replicate the averaging implied by imaging.  Contrasting the
vortex dynamics here suggests that the actual motion of topological defects
through the Fermi gas cannot be properly modelled by the simple \gls{ETF}.  The
agreement seen between~\cite{Ancilotto:2012, *Ancilotto:2012a}
and~\cite{Joseph:2011} thus supports the conclusion that these differences do
not affect bulk dynamical properties.

Coarse graining also adds density to the core of vortices, bringing the density
more closely in line with that of the \gls{SLDA}.  In a proper two-fluid model,
these effects would increase the effective mass of topological defects, for
example, altering their dynamical behaviour as was observed for soliton
dynamics.  \todo{Maybe add a smoothed vortex picture here.}

The degree to which the integrated gap $\int \abs{\Delta}^2$ is conserved
provides a measure of the extent to which one can trust the qualitative results
of the \gls{ETF} model, and Fig.~\ref{fig:DeltaConservation} shows that a
reasonable estimate of this can be obtained from $\int \abs{\bar{\Psi}}^2$ where
$\bar{\Psi}$ is $\Psi$ smoothed on a scale of $q \approx 1.5k_F$ --
i.e. $\bar{\Psi}$ is the result of applying a low-pass filter to $\Psi$
excluding Fourier components with $k > 1.5k_F$. Of course, one can also extract
this information from the spectra (Fig.~\ref{fig:PowerSpectrums}) but the
non-conservation of the integrated gap provides a convenient representation.

To see that this diagnostic applies in other geometries, we consider the success
of~\cite{Ancilotto:2012, *Ancilotto:2012a} where the \gls{ETF} quantitatively
describes the evolution of the densities observed in the
experiment~\cite{Joseph:2011} which collides two clouds of the unitary Fermi
gas.  Checking the diagnostic on one-dimensional and two-dimensional
realizations of this experiment, we find that $\int \abs{\bar{\Psi}}^2$ is
conserved on the sub-percent level, providing evidence that this criterion is
applicable for generic traps, especially where the potential doesn't have large
gradients.

\section{Conclusion}\label{sec:conclusion}%
We study the features and the limitations of the \gls{ETF} as model for the
dynamics of unitary Fermi gases by comparing and contrasting its dynamical
properties with those of the Fermionic \gls{TDSLDA} \gls{DFT}
from~\cite{Bulgac:2011b}.  Like the \gls{GPE}, the dynamical \gls{ETF} model
depends on a single collective wavefunction $\Psi$; it is therefore
significantly easier to solve numerically than the \gls{TDSLDA} which requires
evolving hundreds of thousands of wavefunctions.  Unlike the \gls{TDSLDA},
however, the \gls{ETF} lacks a pair-breaking mechanism.  The extra fermionic
degrees of freedom in the \gls{SLDA} allow it to model both the superfluid and
the normal components whereas the \gls{ETF} models only the superfluid.

By comparing the dynamic response of the \gls{ETF} with that of the
\gls{TDSLDA}, we can assess the importance of these pair-breaking effects on the
overall dynamics.  We find that the \gls{ETF} and \gls{TDSLDA} have similar
properties at low-energies with a similar static response
(Fig.~\ref{fig:static_response}) on the ten-percent level even for momenta $q$
about $2k_F$.  The dynamic linear response (Fig.~\ref{fig:linear_response}) is also
similar for small momenta and frequencies as required by the equation of state,
but significantly departs near the pair-breaking threshold $\omega \sim
2\Delta$. We also remark about a possible physical consequence of the difference
in the curvatures of the phonon dispersion curves, as elucidated by
Fig.~\ref{fig:linear_response}. In a theory with positive curvature, the phonon
with higher energy is kinematically allowed decay to multiple phonons of lower
energy. This is the case in \gls{ETF}, but for $\xi=0.42$, $\eta=0.502$ and
$\alpha=1$ the phonon dispersion in the \gls{SLDA} has a negative curvature.
To see this, however, one must devise sensitive probes as the effect is natually
quite small.

Exciting the \gls{ETF} produces phonons and topological defects (vortices in
this case), but nonlinear excitations amplify high-frequency modes, thereby
creating excitations above the pair-breaking threshold.  Once a system contains
appreciable power above the pair-breaking threshold, detailed dynamics of
topological defects disagree markedly between the two theories -- vortex lattice
crystallize in the \gls{TDSLDA}, for example, but remain chaotic in the
\gls{ETF}. This is expected since topological defects such as domain walls move
with different velocities in the two theories.

Despite these differences in microscopic behaviour, the \gls{ETF} remains a
useful tool for modelling bulk dynamics.  This is perhaps best demonstrated by
the remarkable quantitative agreement between the \gls{ETF}
simulations~\cite{Ancilotto:2012, *Ancilotto:2012a} and the observed shock-wave
phenomena obtained experimentally by colliding two \gls{UFG}
clouds~\cite{Joseph:2011}.  The quantitative agreement conclusively demonstrates
that~\cite{Joseph:2011} is not probing dissipative effects such as viscosity
which are missing in the conservative \gls{ETF} approach.

The picture that emerges is that the high-frequency, short-wavelength phonon gas
in the \gls{ETF} acts very much like the excited ``normal'' component created
through pair-breaking in the \gls{TDSLDA}. This suggests that one might be able
to improve a simple model like the \gls{ETF} by somehow introducing a normal
component and a coarse-graining process that transfers energy and particle
number from the superfluid to this normal component, such as realized for
bosonic systems with the \gls{SGPE} and \gls{SPGPE} (see
Refs.~\cite{Gardiner:2002, *Rooney:2012} and references therein).  This is
further confirmed by coarse graining the \gls{ETF} results which results in a
much better qualitative agreement with the \gls{TDSLDA}.  One effect of coarse
graining is to add density in the cores of topological defects; thus, a theory
that effectively coarse grains should better reproduce the dynamics of fermionic
defects which have a different effective mass.  Another effect is the reduction
in the integrated square of the coarse-grained order parameter, mimicking the
conversion of the superfluid to a normal fluid.  The qualitative agreement of
this reduction also provides a diagnostic for assessing how well a given
\gls{ETF} simulation might model dynamics in the \gls{TDSLDA}.

\begin{acknowledgments}
  \noindent
  We thank A.~Bulgac and A.~Luo for useful discussions and sharing their data. 
  This work is supported, in part, by \textsc{us} \gls{DoE} grant \MMFGRANT.
  and by \gls{NSERC}.
\end{acknowledgments}

\appendix

\section{Numerical Implementation}\label{sec:numer-impl}%
Almost any algorithm implementing the \gls{GPE} can be easily extended to
implement the \gls{ETF}: the only differences are the form of the nonlinear
interaction ($\rho^{5/3}$ vs. $\rho^2$), and a few factors of $2$.  We implement
the evolution using two methods: if high-accuracy is needed (in order to
calculate a power spectrum for example), then we use a fifth-order integrator
described in~\cite{Hamming:1973} that averages the \gls{ABM} predictor-corrector
methods:
\begin{align}
  p_{n+1} &= \frac{y_{n} + y_{n-1}}{2} 
  +\frac{h}{48}\Bigl(
  119y'_{n} - 99 y'_{n-1} + \nonumber\\
  &\qquad+69y'_{n-2} - 17y'_{n-3}\Bigr)
  + \frac{161}{480} h^5 y^{(5)}, \nonumber\\
  m_{n+1} &= p_{n+1} - \frac{161}{170}(p_n - c_n),\nonumber\\
  c_{n+1} &= \frac{y_{n} + y_{n-1}}{2} + \frac{h}{48}\Bigl(
  17m'_{n+1} + 51y'_{n} + \nonumber\\
  &\qquad + 3y'_{n-1} + y'_{n-2}\Bigr) - \frac{9}{480} h^5 y^{(5)}\nonumber\\
  y_{n+1} &= c_{n+1} + \frac{9}{170}(p_{n+1} - c_{n+1}),
\end{align}
where $h = \delta_{t}$ is the time-step.  Here the primes denote derivatives as
computed with the Hamiltonian: $y'_{n} = \partial_{t}y_{n} = -\I\mat{H}y_{n}$.
Each iteration requires two applications of the Hamiltonian -- one for the
predicted step $m'_{n+1}$ (evaluated at time $t + h/2$) and one for the
corrected step $y'_{n+1}$ (evaluated at time $t+h$).  Note that $y_{n+1}$ is
accurate to order $h^6$, so after iterating by $N = T/h$ steps, one obtains an
error that scales as $h^5$ (fifth order). This scaling requires that the
function be at least $\mathcal{C}^{(4)}$, so a high-order integrator must be
used to provide the first four stating iterations.  (Another approach is to
start from a stationary solution so that $y_{m} = p_{m} = c_{m} = y_{0}$ for
$n\in \{1,2,3\}$.

This method requires storing the previous four derivatives $y'_{n-3}$,
$y'_{n-2}$, $y'_{n-1}$, $y'_{n}$, as well as the previous $p_{n} - c_{n}$ and
current and previous steps $y_{n}$, and $y_{n+1}$. Adding an additional
workspace for computing the \gls{FFT}, the memory requirements rise to $8N$
complex numbers.

When one does not need high accuracy, an alternative method,
the split-operator approach, is faster.  Here one decomposes the Hamiltonian
into kinetic and potential parts, each of which can be applied directly to the
wavefunction with an error that scales as $h^3$~\cite{Huyghebaert:1990}:
\begin{gather}
  e^{\I\hbar h (\mat{K} + \mat{V})} 
  = e^{\I\hbar h \mat{K}/2}e^{\I\hbar h \mat{V}}e^{\I\hbar h \mat{K}/2} + \order(h^3).
\end{gather}
This method is symplectic, effecting strictly unitary evolution, and requires no
additional storage beyond the current state and any scratch space needed for
computing the \gls{FFT}.  In addition, this approach can be nicely transferred
to a \gls{GPU} for a further gain in performance. Although not as accurate as
the higher-order \gls{ABM} method, this method can be used with relatively large
time steps ($h = \delta_t \approx 0.1/\hbar E_c$) to quickly gain a qualitative
picture of the dynamics.

\section{Parameters of the trap}\label{sec:trap-params}%
To compare our results with Ref.~\cite{Bulgac:2011b} we use $\xi=0.42$ in
Eq.~\ref{eq:ETF}.  The parameters are conveniently written in ``atomic'' units,
where we take $\hbar=m=1$. The chemical potential chosen so that the ground
state density of fermions, $\rho$, at the centre of the trap matches the desired
value, $\rho_{\text{central}}=0.0375$ which corresponds to $k_F=1.035$
($E_F=0.536$).  This fixes the chemical potential $\mu=\xi E_F=0.225$.

The trapping potential is cylindrically symmetric and taken to have the same
profile as used in~\cite{Bulgac:2011b}.
\begin{equation}
  V(r) = 3.9478\times\left[\frac{
  1-\cos\frac{2\pi r}{L}}{2}\right]^{8}
\label{eq:trap}
\end{equation}
where $L=32$ is the extent of the simulation box in each direction, and $r$ is 
the distance from the centre. 

The trap radius $R$ defined as the point where $V(r/0.90)=\mu$, where the
factor of $0.9$ is used to avoid the periphery of the cloud. This gives
$R=9.08$.  We focus on the family of simulations performed
in~\cite{Bulgac:2011b} for $N=32$ grid points in each ($x$ and $y$) direction.
To compare, we perform simulations for $N=32$ and $N=64$ points in each
direction. In table~\ref{tab:vortices_vs_vAurel}, the trap has radius $R=9.08$
and the stirrer orbits at fixed distance $r_{\text{stir}} = 6$ from the center.

We begin the simulation with the ground state in the potential~\ref{eq:trap} at
time $t=0$. If the Thomas-Fermi profile were exact in both \gls{SLDA} and 
the \gls{GPE}, the densities profile the two would be equal. In reality, the
density profiles differ near the boundary of the trap and the total number of
particles in the \gls{GPE} ($N_{\text{part}}=9.047$ per unit length, or $289.5$
particles in a cylinder of height $32$) differs slightly from the total number
in the \gls{SLDA} ($N_{\text{part}}=9.375$ per unit length, or $300$
particles in a cylinder of height $32$).

A stirring potential of the form,
\begin{equation}
  V_{\text{stir}} = E_F \exp\left(-\frac{r^2}{r^2_{\text{pin}}}\right)
\end{equation}
with $r_{\text{pin}}=2$ is gradually switched on after $t=94.25/e_F$ and
switched off after stirring the superfluid $n_{\text{stir}}$ times.

\exclude{
\begin{figure}[tb]
  \includegraphics[width=\columnwidth]{force_vs_t}
  \caption{\label{fig:force}(color online) % 
    Forces for the simulations for $v_{\text{stir}}=0.1 v_F$,
    ($v_{\text{stir}}=0.11 v_F$) $v_{\text{stir}}=0.2 v_F$,
    ($v_{\text{stir}}=0.197 v_F$), and $v_{\text{stir}}=0.25 v_F$
    ($v_{\text{stir}}=0.242 v_F$) for \gls{TDSLDA} (\gls{ETF}). }
\end{figure}
}

\section{Energetics}
In Fig.~\ref{fig:E_vs_t} we compare how the total energy of the system changes
through four sample stirring simulations. 

Both \gls{ETF} and \gls{TDSLDA} simulations with $v_{\text{stir}} \approx
0.1v_F$ that result in a single vortex display the same qualitative behaviour:
First the energy increases as the stirring potential is turned on and fluid is
displaced (slight quantitative differences on the 10\% level appear here due to
the aforementioned differences in the displaced densities).  The stirring
potential nucleates a vortex from the edge of the trap, and then effectively
pins the vortex: the stirring potential displaces fluid, thereby creating an
attraction for the vortex which also prefers a density depletion in its core.
The vortex then oscillates in this pinning potential, causing an oscillating
force on the stirring potential that appears as oscillations in the energy
$\d{E}/\d{t} = -\vect{F}\cdot\vect{v}$ as the stirring potential does work on
the system.  Since usually only one vortex is attached to the pinning site at
any given time, the $\vect{F}$ can be associated with the pinning force exerted
by the pinning potential on the vortex. Since the velocity of the stirrer,
$\vect{v}$, is almost identical in the two simulations, this points to the fact
that $\vect{F}$ (per unit length) is larger in the \gls{GPE} compared to
\gls{SLDA}. This is not surprising because the force exerted on the superfluid
by a potential $V$ can be written as,
\begin{equation}
  \vect{F} = -\int d^2  x \rho\vect{\nabla} V
\end{equation}
and the depletion in the density in the core of the vortex which is the source
of the pinning force, is smaller in the \gls{SLDA} compared to the
\gls{GPE}.
% It is interesting that the depletion in the density caused by the potential is
% more pronounced in the SLDA.

Finally, the stirring potential is removed, leaving the single vortex, which an
orbit determined by a counter-circulating image vortex outside the
trap~\cite{PS:2002}.  The subsequent motion of the vortex is almost identical
in both \gls{ETF} and \gls{TDSLDA} simulations since it results from
long-distance superfluid hydrodynamic boundary effects rather than buoyant
force (the trap is flat at the orbital radius) which would be more sensitive to
the difference in vortex mass due to non-zero occupation of the core in the
\gls{TDSLDA}. 

For $v_{\text{stir}}\approx 0.2v_F$, a similar picture is presented: vortices
are nucleated from the boundary of the system, and one remains pinned to the
stirrer while three others perform complex orbits as governed by the Magnus
relation in the presence of each other, the boundary of the trap, and the
stirrer. The motions of the vortices appears to be chaotic -- small changes in
initial conditions, lattice resolution, etc. lead to different
trajectories\todo{Figure???}.  For example, in the high-resolution $64^2$
\gls{ETF} simulation with $v_{\text{stir}} = 0.2v_F$, eventually four vortices
remain in the bulk, whereas with the $32^2$ simulation and the $v_{\text{stir}}
= 0.197v_F$ simulations, one vortex attaches itself to the boundary of the trap
and vanishes once the stirrer is removed, as in the corresponding \gls{TDSLDA}
simulation.  At several times during the simulation, the stirring potential
catches up with one of the free vortices and the vortex-pinning interaction
exerts a stronger force on the stirrer, allowing it to perform work on this
system: this appears as jumps in the energy evolution.

For $v_{\text{stir}}=0.25v_F$, the \gls{ETF} and the \gls{TDSLDA} simulations
have both qualitative and quantitative differences.  The stirrer in the
\gls{TDSLDA} leaves five vortices in the bulk of the trap, while the $32^2$
\gls{ETF} simulation ends with only two vortices.  Increasing the resolution to
$64^2$ gives six final vortices, though for a slightly smaller stirring speed
($v_{\text{stir}}=0.242$) five vortices are left, matching \gls{TDSLDA}.  In the \gls{ETF}, however, the stirrer
creates filamentary structures that are not seen in the \gls{TDSLDA}.  For
higher $v_{\text{stir}}$, the qualitative differences between the two models
become even more pronounced: in particular, as discussed before, the
low-resolution \gls{ETF} simulations become so noisy that it is difficult to
identify the vortices.  With higher resolution, however, the \gls{ETF} retains
the somewhat striking property of the fermionic simulation, that coherent
superfluidity persists for super-sonic stirring $v_{\text{stir}} > c_s \approx
0.37v_F$. As pointed out in~\cite{Bulgac:2011b}, this is due to the compressible
nature of the superfluid: the stirring potential compresses the superfluid,
raising the local critical velocity.  Of course, strictly speaking, the
\gls{ETF} always remains superfluid, but for sufficiently fast stirring, the
phase fluctuations are so rapid that a coarse-grained picture will find little
spatial coherence. An improved two-fluid model would include such a
coarse-graining procedure to convert superfluid to the normal fluid.

\begin{figure}
  \includegraphics[width=\columnwidth]{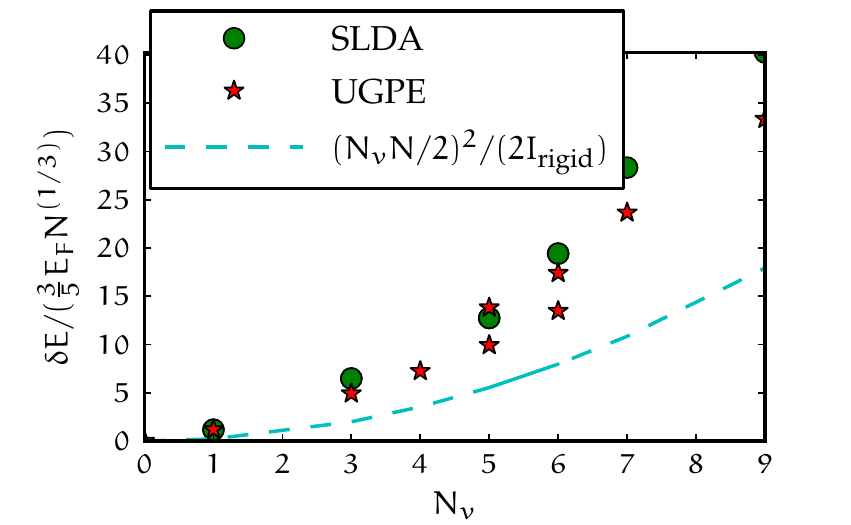}
  \caption{\label{fig:E_excit}%
    The excitation energy (the difference between the energy after the removal
    of the potential and before the introduction of the potential) as a function
    of the number of vortices created in the bulk ($N_v$).}
\end{figure}

From the total energy difference after the stirrer is removed we can calculate
the total excitation energy.  This depends primarily on the number vortices
added to the system, though there are minor contributions due to excited
phonons. The classical estimate~\cite{Bulgac:2011b}
\begin{equation}
  E_{\text{excit}} \sim L_z^2/(2 I_{\text{rigid}}),
  \label{eq:Eexcit_classical}
\end{equation}
where $L_z$ is the angular momentum which is roughly proportional to the number
of vortices, and $I_{\text{rigid}}$ is the moment of inertia of the superfluid
in the trap, suggests that it increases quadratically with the number of
vortices. Eq.~\ref{eq:Eexcit_classical} just a rough estimate, and it turns out
to overestimate the excitation energy by a factor of two, as we show in
Fig.~\ref{fig:E_excit}.  The \gls{GPE} provides a reasonable estimate of the
excitation energy in the \gls{SLDA} as a function of the number of vortices
created, all the way up to $v_{\text{stir}}<v_F0.35$.

\clearpage
\bibliographystyle{apsrev4-1}
%\bibliography{local,master}
\end{document}